\documentclass[runningheads]{llncs}
\usepackage{graphicx}
\usepackage{listings}
\usepackage{xcolor}
\usepackage{float}

\begin{document}
\title{An approach to performance portability through generic programming}
%
%
\author{Andreas Hadjigeorgiou\inst{1} \and
Christodoulos Stylianou\inst{2} \and
Mich\`{e}le Weiland\inst{2} \and
Dirk Jacob Verschuur\inst{3} \and
Jacob Finkenrath\inst{1}}
\authorrunning{Hadjigeorgiou A. et al.}
%
\institute{The Cyprus Institute, Nicosia, Cyprus \\ 
\email{a.hadjigeorgiou@cyi.ac.cy} \\
\email{j.finkenrath@cyi.ac.cy} \\
\and EPCC, University of Edinburgh, Edinburgh, UK \\ 
\email{c.stylianou@ed.ac.uk} \\
\email{m.weiland@epcc.ed.ac.uk} 
\and Delft University of Technology, Delft, The Netherlands \\ \email{d.j.verschuur@tudelft.nl}
}
\maketitle              
\begin{abstract}

The expanding hardware diversity in high performance computing adds enormous complexity to scientific software development. Developers who aim to write maintainable software have two options: 1)~To use a so-called data locality abstraction that handles portability internally, thereby, performance-productivity becomes a trade off. Such abstractions usually come in the form of libraries, domain-specific languages, and run-time systems. 2)~To use generic programming where performance, productivity and portability are subject to software design. In the direction of the second, this work describes a design approach that allows the integration of low-level and verbose programming tools into high-level generic algorithms based on template meta-programming in \texttt{C++}. This enables the development of performance-portable applications targeting host-device computer architectures, such as CPUs and GPUs. With a suitable design in place, the extensibility of generic algorithms to new hardware becomes a well defined procedure that can be developed in isolation from other parts of the code. That allows scientific software to be maintainable and efficient in a period of diversifying hardware in HPC. As proof of concept, a finite-difference modelling algorithm for the acoustic wave equation is developed and benchmarked using roofline model analysis on Intel Xeon Gold 6248 CPU, Nvidia Tesla V100 GPU, and AMD MI100 GPU.

\end{abstract}
\section{Introduction} \label{intro}

The diversity of computer architectures in High Performance Computing (HPC) has evolved remarkably in the past two decades. During this period, HPC systems advanced from {\itshape single-core} processing units (PUs), to {\itshape many-core} PUs. In other words, performance scaling has been achieved, to a big extend, with additional parallelism at PU-level. However, parallelism comes in different forms. Briefly, CPUs and GPUs dominate the HPC landscape today. At the same time, even PUs that fall within the same category, e.g. CPU, may pose differences in their architecture based on how cores are physically placed on chips, e.g. NUMA regions, etc. This evolution has dramatic consequences on the software development side. A single optimized C or FORTRAN code that has been considered a performance portable solution in the past is neither sufficient nor fully-portable at all nowadays. As a result, during the past decade efforts were made for the development of abstraction layers, such as Kokkos \cite{kokkos_i} and RAJA \cite{raja}, that enhance the development of portable HPC applications. At the same time, these efforts aim to increase productivity, by easing the development effort. The outcomes from these efforts come in the form of programming models that provide a set of \textit{memory spaces}, \textit{execution spaces} \& \textit{policies}, \textit{iteration ranges}, \textit{data layouts}, and other concepts, which serve as building blocks for the development of parallel \textit{single-source} code that is performance portable to one degree or another \cite{perf_port_across}.

Performance critical software is always developed using low-level verbose programming tools, which expose to the developers ways to optimally map code to the target hardware \cite{gometa}. On top of this, performance optimization usually requires considerable changes in the software design. Therefore, performance should be considered in the design phase of the software, a step that is usually underestimated in computational science \cite{SEforCS,SoftArch}. This work is not an attempt to provide an alternative programming model equivalent to the performance portability layers described in the previous paragraph, despite the fact that it is inspired from them. Rather, we discuss a design approach that allows developers to achieve performance and portability by separating two concerns: the algorithm as a multi-step process that evolves in-time, and its actual code implementation details. This separation is achieved through static polymorphism using meta-programming techniques based on \texttt{C++} templates, which falls under the broad category of \textit{generic programming} \cite{meta}.

\begin{figure}
    \centering
    \includegraphics[width=0.7\textwidth]{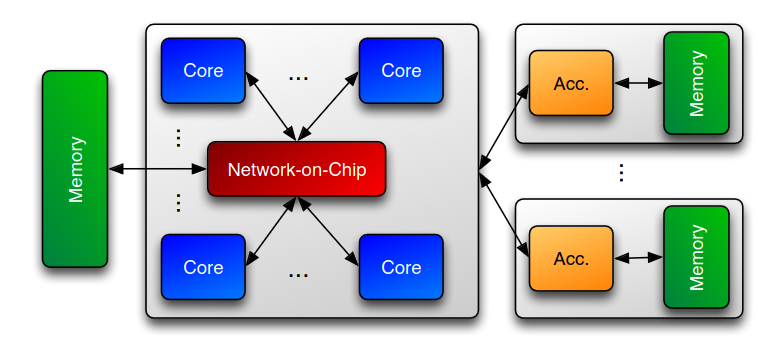}
    \caption{Host-device computer architecture as an {\itshape abstract machine model} \cite{host_device}.}
    \label{host_device}
\end{figure}

The host-device model introduced as an {\itshape abstract machine model} towards exa-scale computing \cite{host_device}, refers to a machine that has a multi-core processor (host), which is coupled with one or more discrete accelerators (devices).  Each accelerator has a local, high throughput, and physically separated memory as illustrated in Figure \ref{host_device}. The model represents accurately the node architecture of most supercomputing systems nowadays. The discussed design approach allows to integrate low-level programming techniques into high-level abstractions in order to target in a performance portable manner host-device computer architectures.

By no means we support this is a silver bullet approach to solve the performance portability challenge, or that alternative approaches might not be proven equally well. It is however, a robust and simple approach to achieve performance and portability, and good software engineering practises to be incorporated in the development process, such as: separation of concerns, extensibility, ease of testing, and memory safety. As a proof of concept (POC), a two-dimensional Finite-Difference (FD) modelling scheme for the acoustic wave propagation is developed and made publicly available\footnote{\url{https://github.com/ahadji05/pp-template/}} on GitHub. We select this POC application since its stencil-based nature represents well enough a range of existing applications and at the same time it is simple and broadly understood by the general audience of HPC.

\section{Performance portability by design}

From the science point of view, code is seen as an asset that allows scientists to do more in terms of simulations, analyses, visualization, etc. However, from the software engineering point of view, code is debt. The larger the code-base it is, the more difficult is to manage. Software design, is what allows it to be manageable and grow without collapsing from its own weight. These concepts have been of great concern in the broad computer science domain since long time ago \cite{design}. For scientific computing, the challenge introduced from the emerging hardware diversity in HPC makes this necessity even more apparent. To our experience, the major problem in many scientific applications is usually not the effort itself that is needed for re-writing some part of the code for portability to another architecture. It is rather that by design the applications have architecture-specific code to all their extent and silent assumptions about the location of data, thus, the transition becomes a redundant, not well defined, and as a consequence erroneous procedure. All these problems, become even more apparent due to the lack of a regression test framework in most cases. Developing applications for heterogeneous computer architectures requires radical rethinking of the scientific software development approach.

For the development of HPC applications we came across two approaches. On the one hand, is the use of so-called data locality abstractions \cite{trends} (DLAs) that come in the form of libraries, domain-specific languages, or run-time systems with common theme the increase of productivity. Portability is handled internally from the DLAs, thus, the developer does not require hardware specific knowledge. On the other hand, is the use of meta-programming techniques \cite{gometa} that allow integrating verbose programming tools into generic interfaces with common theme the increase of performance. Portability is expanded by integrating more programming models under the generic interface. In practice, portability is easier following the first approach, however, the performance boundary can be pushed further following the second one. The approach described in this work is in the direction of the second. Using \texttt{C++} templates and meta-programming techniques, we discuss the development of generic containers, routines, and algorithms. The term generic denotes \textit{template-with-respect-to} two types/concepts: Memory-Space and Execution-Space. These serve as the core types that allow to form a generic approach to target host-device computer architectures.

\begin{figure}[t]
    \centering
    \includegraphics[width=0.8\textwidth]{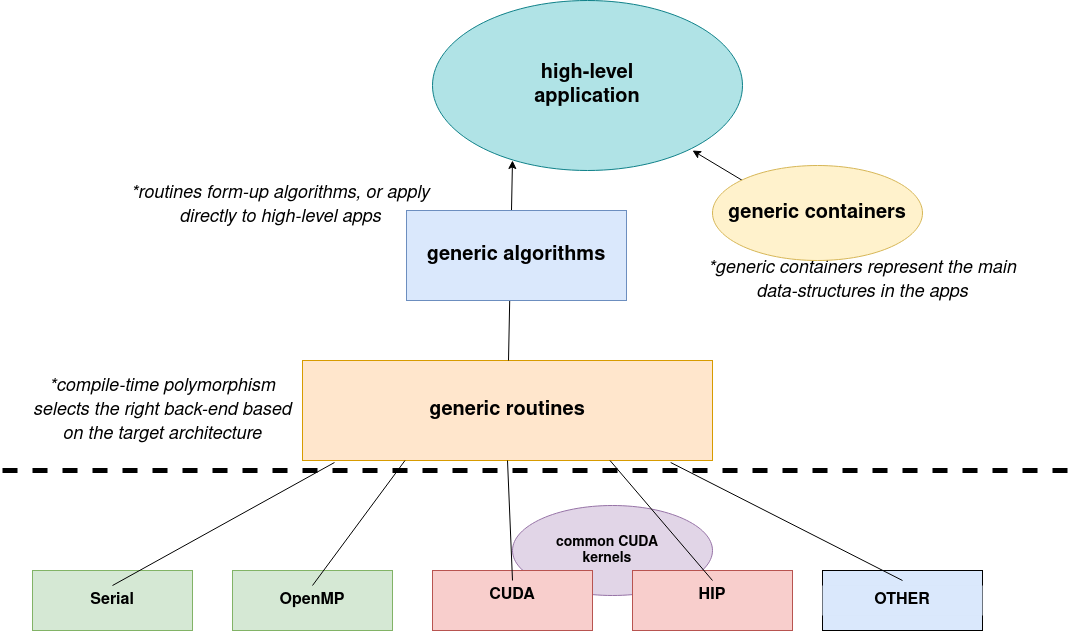}
    \caption{Software design for performance and portability through generic programming.}
    \label{fig:code_design}
\end{figure}

Figure \ref{fig:code_design} illustrates the software design that we describe in this work. The dashed horizontal line demonstrates the separation of two concerns: 1) the development of generic building-blocks, 2) the implementation details. In this case, generic denotes architecture-agnostic, whereas details denotes architecture-specific. The end goal is to develop building-blocks in the form of \textit{containers}, \textit{routines}, and \textit{algorithms}, which allow the development of applications that are transferable across architectures. Then, based on compile-time choice, the behaviour of the building-blocks is resolved according to the implementation details we select. Finally, the behaviour has two aspects: first, the location of the data, second, the access pattern.

This design assumes that host is always a CPU that is programmable with \texttt{C++}, and the device could be either the host itself, or an accelerator programmable with a low-level programming tool that is supported in the back-end. In our case the back-end is developed using \texttt{OpenMP}, \texttt{CUDA} and \texttt{HIP}. \texttt{Serial} is the sequential processing back-end that serves as reference for testing. At the same time, the application is extensible to any other back-end we may want to develop in the future. As long as a future implementation adheres to the interface that is already in-place, all code on top of is reused without any additional change. 

\section{Concern 1: Building blocks} \label{concernI}

The core concepts we identify as minimum requirement for separating the implementations details from the generic algorithms are two: Memory Space and Execution Space. The role of Memory Space is to define the location of the data, whereas the Execution Space selects the right back-end that provides the implementation details. In fact, this allows to leverage the type system of the language in order to develop compile-time rules that impose memory safety.

\begin{figure}
    \centering
    \includegraphics[width=0.8\textwidth, height=2.5cm]{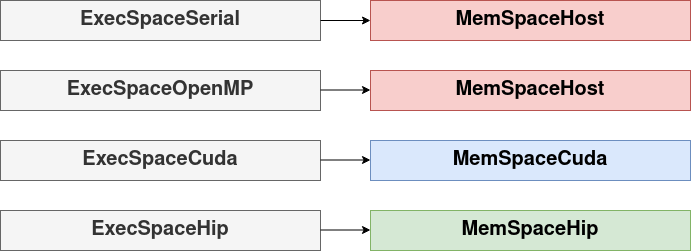}
    \caption{Execution spaces and their corresponding Memory Spaces for targeting CPU and GPU architectures.}
    \label{fig:core}
\end{figure}

The Memory Spaces are concrete classes that provide five basic memory management operations: \texttt{allocate}, \texttt{release}, \texttt{copy}, \texttt{copyFromHost}, and \texttt{copyToHost}. These operations are implemented as static methods so they are bound to their class name. Thus, a Memory Space passed as template parameter to other generic classes provides these operations. The right hand side of Figure \ref{fig:core}, illustrates the three Memory Spaces proposed and developed in our case. The default is the host memory space, namely \texttt{MemSpaceHost}. For this particular space, the three copy operations have the same underlying implementation; copy data within host memory space. They diverge only for the other two memory spaces \texttt{MemSpaceCuda} and \texttt{MemSpaceHip}. On the left hand side in Figure~\ref{fig:core}, the Execution Spaces, are concrete classes as well that are being used for explicit specialization of routines based on the \textit{tag-dispatching} idiom that is discussed in Section~\ref{concernII}. Moreover, each Execution Space has one type-trait that defines the accessible Memory Space. This allows the development of type-rules to ensure that the accessible data are on the right location.

\noindent For the development of architecture-agnostic applications we identify three generic building blocks: containers, routines, and algorithms. They are generic in the sense that their type is \textit{to-be-specified} according to template parameters. The two template parameters that define them are: Memory Space and Execution Space. In essence, the {containers} hold the data that are processed during run-time in a suitable location, processing comes from the {routines} that act on the containers with a suitable access pattern, and finally, the {algorithms} are collections of routines and containers.

\medskip

\noindent \textbf{Containers} are generic \texttt{C++} data-structures, e.g. class or struct, with one template parameter that resolves to a valid Memory Space. The scope of the Memory Space is to determine the location of the container's data by providing the five basic memory management operations that we mentioned earlier. Containers represent the main entities in our application. For example, for the finite-difference scheme that we developed as POC app, we developed the \texttt{ScalarField<Mem>} container to represent velocity and wave fields. Containers may have metadata explicitly on host that serve control-flow and run-time assertions. The data related to the processing done during the simulation are located on the provided Memory Space (see template parameter \texttt{Mem}) so they are directly accessible from the routines. 

\medskip

\noindent \textbf{Routines} are generic \texttt{C++} functions with two template parameters: An Execution and a Memory space. The routines operate on the containers and provide the processing steps that our application involves. The Memory Space determines the location of data in the containers that are passed as parameters to the function. The Execution Space is used as a tag parameter that dispatches the routine to a back-end implementation. This approach leverages the function overloading feature of \texttt{C++}. The Execution Space and the Memory Space need to be compatible, otherwise the application could crash due to an invalid memory access. The language's type system allows to develop rules, e.g. using \texttt{std::enable\_if}, in order to let the compiler inspect this compatibility. This allows to capture with a meaningful message an invalid implementation at compile-time, instead of having to debug an invalid memory access that occurs out of the blue.

\medskip

\noindent \textbf{Algorithms} are generic \texttt{C++} classes with a single template parameter representing the Execution Space. The Execution Space provides its accessible Memory Space as a type-trait, which is used internally in the body of the class. The algorithms are collections of containers and routines in the sense that they have containers as member variables and their methods apply operations on them through one or more routines. The class provides no information whether and how the computation is parallelised, and if data reside on the CPU or the GPU, or so. All these concerns resolve to the implementation that is a separate part of the code. The scope of the algorithm as a meta-program is first, to make sure that given an Execution Space, the correct Memory Space is selected based on which the containers allocate their data suitably. Second, based on the Memory Space and the Execution Space the routines dispatch at compile-time to the correct back-end implementation. This design approach adds to the quality of the code in the sense that it makes it more easy to read and understand, reusable across different architectures due to its generic nature, and memory safe. At the same time, since the implementation is decoupled from the interface, we can use any low-level programming tool to develop it efficiently based on the target-hardware without altering the aforementioned.

\section{Concern 2: Implementation details} \label{concernII}

In Section~\ref{concernI}, we discussed the development of generic containers, routines, and algorithms based on two template parameters: Memory Space and Execution Space. These serve as the building blocks for the development of performance portable applications. In here, we discuss how we develop the architecture-specific implementations in such way that they adhere to the common abstract interface and can be identified at compile-time through tag dispatching. 

As an example, let us consider a generic routine, namely \texttt{dosomething}, that applies some processing on a container \texttt{X}. The specializations\footnote{In \texttt{C++} this is so-called explicit (full) template specialization} for \texttt{OpenMP} and \texttt{CUDA} implementations are shown in Listings~\ref{lst:omp_impl} and~\ref{lst:cu_impl} respectively.
\begin{lstlisting}[language=C++,caption={Specialisation of the OpenMP implementation.}, label=lst:omp_impl, captionpos=b]
template<>
void dosomething(X<MemSpaceHost>& A,... ,ExecSpaceOpenMP tag){
    \\ parallel OpenMP code ...
}
\end{lstlisting}
\begin{lstlisting}[language=C++,caption={Specialisation of the Cuda implementation}, label=lst:cu_impl, captionpos=b]
template<>
void dosomething(X<MemSpaceCuda>& A,... ,ExecSpaceCuda tag){
    \\ config-launch CUDA kernel ...
}
\end{lstlisting}

\noindent Listing~\ref{lst:omp_impl}, indicates the function definition for the \texttt{OpenMP} back-end. This specialization, receives as parameter, among others, a container \texttt{X} whose data is allocated on host using the corresponding Memory Space. Thus, the data are accessible by a parallel \texttt{OpenMP} implementation. In Listing~\ref{lst:cu_impl}, the function is launching a \texttt{CUDA} kernel, thus, container needs to have the data on device. In that case, this is guaranteed by \texttt{MemSpaceCuda}, which is the Memory Space of container \texttt{X}.

Each template specialization is implemented in a different \textit{translation-unit} isolated from others. Alternatively, conditional guards\footnote{\texttt{\#ifdef}, \texttt{\#else}, \texttt{\#endif}, etc.} are used to diverge source-code compilation based on which architecture is targeted. More specializations can be developed as long as they adhere to the interface that is provided from the generic routine, and a valid combination of Memory-Execution spaces is used. In that sense, a generic codebase is extensible to other hardware without conflicts with existing parts of the code. As a result, extending portability becomes a well defined procedure that is trivial and memory-safe. At the same time, performance is not compromised or limited by any third party library since it can be specifically implemented with any low-level programming tool of choice.

Based on the design approach that is described in this work, we developed four Execution Space options and three Memory Space options as shown in Figure \ref{fig:core}. The \texttt{Serial} and \texttt{OpenMP} Execution Spaces are compatible with the \texttt{MemSpaceHost}. The \texttt{CUDA} and \texttt{HIP} Execution Spaces are compatible with the \texttt{MemSpaceCuda} and \texttt{MemSpaceHip} respectively. These options are sufficient to cover portability on the majority of HPC systems as of today.

\section{Start and stop} \label{startstop}

The discussed design approach targets host-device computer architectures, thus, the start and stop of applications need to pass through the host always for I/O purposes. Thereby, in a program's lifetime, there must coexist one explicit host Memory Space, namely \texttt{MemSpaceHost}, and one alias for device Memory Space, namely \texttt{memo\_space}. By default, \texttt{memo\_space} can be the host Memory Space itself if no accelerator is targeted.

\smallskip

\noindent \textbf{How an application starts:} Applications are organized based on the assumption that there is a host PU whose resources are managed with \texttt{MemSpaceHost}, and a device whose resources are managed with \texttt{memo\_space}. The latter, is defined at compile-time based on the target architecture. All input data are allocated using \texttt{MemSpaceHost}, so they are initialized on host. All containers are instantiated using \texttt{memo\_space}. The, input data are copied into the containers from host using the method \texttt{memo\_space::copyFromHost}. Once the data are copied into the containers they are directly accessible from the generic routines and algorithms. This approach fits the host-device model, which was introduced in Figure~\ref{host_device}, because it is a unified approach that allows to initialize the generic containers with data, either on host, or the device, using the same interface.

\smallskip

\noindent \textbf{How an application ends:} An application ends with output data printed on screen and/or stored in output files. To meet this requirement within the context of the host-device model we follow an analogous to the previous paragraph's approach. The containers have their data managed by \texttt{memo\_space}, which resolves either to host, or the device memory. Thereby, before output there must be an explicit call for transfer to the host memory. In analogoy to the previous paragraph, this is performed via the method \texttt{memo\_space::copyToHost}.

\section{Testing as an integral part}

Testing is the proof of correctness and should be an integral part of scientific software. The purpose is to verify in a quantitative manner that the implementation is correct. It applies to individual routines such they are expected to return a specific output given a specific input, so-called \textit{unit-tests}. Additionally, software should be tested in the connectivity between different parts as error may exist in the glue-code, namely \textit{integration-tests}. Testing increases the development curve, however, the returns pay-off the effort and provide a proof that indeed the software behaves exactly as it should. According to Prabhu et al. \cite{Prabhu}, scientists spend as much as half of their programming effort on finding and fixing errors using "primitive" debugging approaches.

Within the context of our discussion, the development of implementations using different programming tools that provide the same routines and algorithms make the need for testing even more apparent. Ideally, we do not want the effort of developing different architecture-specific implementations to extend the testing effort. This is possible, if the tests are applied to the layer of generic containers, routines, and algorithms. To do so, the tests are developed based on the ideas that we discussed in Section~\ref{startstop}. We can summarize the procedure in the following steps:

\begin{enumerate}
    \item the input data are initialised explicitly on Host (\texttt{MemSpaceHost})
    \item using \texttt{memo\_space::copyFromHost} the input data are copied into containers
    \item the routine(s) under testing are invoked
    \item using \texttt{memo\_space::copyToHost} the output data are copied back to Host
    \item test-assertions are performed on Host \ .
\end{enumerate}

The \texttt{memo\_space} is an alias that by conditional compilation resolves to one of the available Memory Spaces from Figure \ref{fig:core}, depending on which backend we want to test against. Based on this approach, all implementations are tested through the same input-output criteria and verify their equivalence. Both unit and integration tests can be effectively implemented based on this approach. Furthermore, this makes code easier and safer to extend in the sense that if a new hardware emerges and we want to develop an implementation to target it, the tests that are already in-place serve as the channels to "pass" through.

\section{Proof of concept}

As POC, a finite-difference modelling scheme for the two-dimensional acoustic wave equation:
\begin{equation}
    \label{eq:acoustic_wave}
    \frac{\partial^2 P(t,x,z)}{\partial t^2} = v(x,z)^2 \bigg( \frac{\partial^2 P(t,x,z)}{\partial x^2} + \frac{\partial^2 P(t,x,z)}{\partial z^2} \bigg) + S(t) \ ,
\end{equation} 
has been developed, where $P$ is the pressure amplitude, $v$ is the space-dependent velocity, and $S$ is a time-dependent source. The FD modelling is performed by the \texttt{WaveSimulator} algorithm\footnote{https://github.com/ahadji05/pp-template/tree/main/include/algorithms}, which has been developed based on the discussion for generic building-blocks in Section~\ref{concernI}. Listing~\ref{lst:wave_sim} is a brief view of the algorithm's interface that shows how the containers and routines that form up the algorithm come together.
\begin{lstlisting}[language=C++,caption={Generic interface for the Wave simulation algorithm.}, label=lst:wave_sim, captionpos=b]
template<class ExecSpace>
class WaveSimulator {
  public:
    using MemSpace = typename ExecSpace::accessible_space;
    void run(){
      add_source(ScalarField<MemSpace>&P, ..., ExecSpace());
      fd_pzz(ScalarField<MemSpace>&Pzz, ..., ExecSpace());
      fd_pxx(ScalarField<MemSpace>&Pxx, ..., ExecSpace());
      fd_time(ScalarField<MemSpace>&Pnew, ..., ExecSpace());
      swap(Pold, P);
      swap(P, Pnew);
    }
    //...other public methods e.g. set/get
  private:
    ScalarField<MemSpace> Pnew, P, Pold, Pxx, Pzz, V;
    d_type _dt, _dh;
    //...other private variables and methods
};
\end{lstlisting}
Initially, \texttt{MemSpace} is the Memory Space that is defined as type-trait from the provided template parameter \texttt{ExecSpace}. The member variables Pnew, P, etc. that store wavefields as well the velocity model V are represented by the generic container \texttt{ScalarField}, which is defined based on the template parameter \texttt{MemSpace}. The algorithm's steps are ordered in the body of the main function, namely \texttt{run()}, which invokes the routines that compose it, namely, 1)~\texttt{add\_source}, 2)~\texttt{fd\_pzz}, 3)~\texttt{fd\_pxx}, and 4)~\texttt{fd\_time}. The algorithm as a meta-program is completely agnostic with respect to any target hardware because both the location of the data as well the implementation details are resolved by the template parameters \texttt{MemSpace} and \texttt{ExecSpace} respectively.
\vspace{-4.5mm}
\begin{figure}
    \centering
    \includegraphics[width=0.5\textwidth]{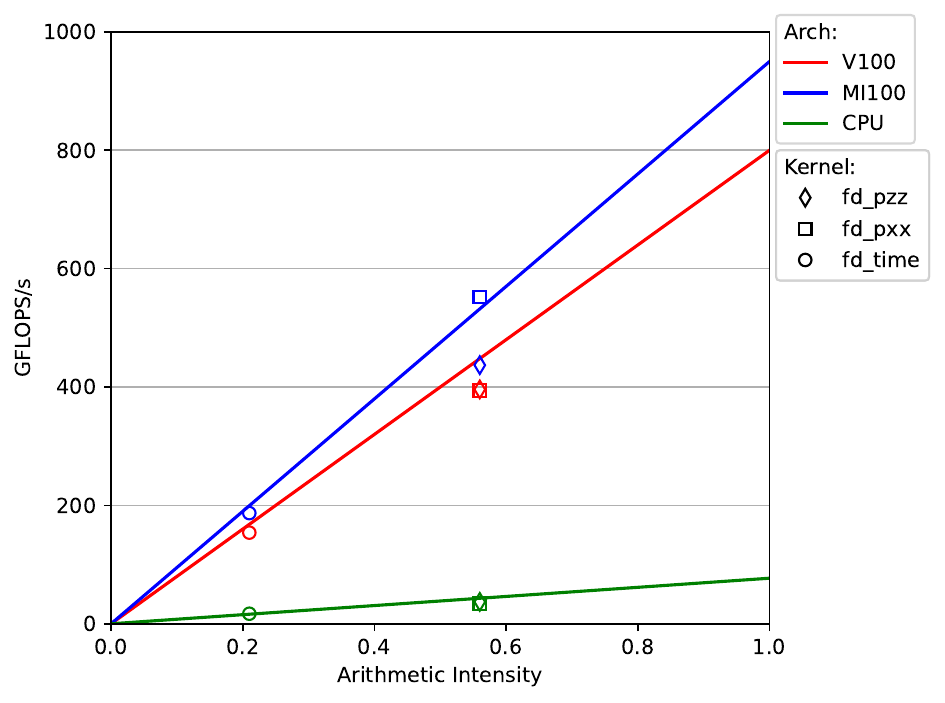}
    \setlength{\belowcaptionskip}{-10pt}  
    \caption{Roofline-model analysis of individual kernels on CPU and GPU architectures.}
    \label{fig:roof}
\end{figure}

Since the implementation details can be developed with any programming model, language, or library of choice we can use a suitable tool for each architecture we target. We use \texttt{OpenMP} for threaded parallelism on CPUs, \texttt{CUDA} for Nvidia GPUs, and \texttt{HIP} for porting \texttt{CUDA}-kernels to AMD GPUs. In Figure \ref{fig:roof}, we demonstrate by roofline model analysis that we achieve near optimal performance in all kernels\footnote{the \texttt{add\_source} kernels has no parallelism to exploit, thus, it is neglected.} of interest. The architectures that we used for this experiment are: Nvidia Tesla V100 GPU, AMD MI100 GPU, and Intel Xeon Gold 6248 CPU. For each case, the practical main-memory bandwidth is measured using the open-source BabelStream benchmark \cite{babel}, and plotted on the figure with a different colour. The curve with the highest slope has the highest memory throughput, which in that case is the AMD MI100 GPU. Based on the arithmetic intensity (Flops/Byte), and the achieved performance (GFlops/s), each kernel is marked on the plot. The colour indicates architecture and the symbol indicates the kernel-name. The evaluation shows that all kernels, except from \texttt{fd\_pzz} on the MI100 GPU, achieve higher than 90 \% efficiency. 

\section{Discussion}

This design approach is powerful because the generic behaviour can be developed without any dependency on external libraries. A C++11 or above compliant compiler is all that is needed. The implementation details that provide the architecture-specific libraries or programming models are needed partially depending on the target hardware. Extensibility is one of the virtues of the described design approach because targeting new hardware becomes a well defined procedure. The first step is to develop the Memory Space that provides the resource management operations. The, second is to develop the Execution Space that is used for tag-dispatching. Once these two classes are developed the back-end for each routine can be implemented in isolation from the others based on two requirements: 1) adhere to the interface, and 2) pass the generic unit-test that is already in place. This strategy allows to implement back-end for a new hardware, e.g. FPGA, or even specialize for specific instruction sets, e.g. AVX, SVE, etc. Furthermore, we argue that productivity does not necessarily come only with the ability to develop code faster. It comes with good separation of concerns so that processes are dissected in small, well defined, understandable pieces that can be developed independently. If a software design is in place and it is understood from the people who develop the different back-ends, productivity is achieved in the most effective way. This work aligns towards this direction. Finally, the discussed design approach is used for the development of preparatory access software within the Delphi Consortium for applications related to seismic wavefield modelling, imaging, and inversion. Similar design approaches were successfully applied in the area of Sparse Linear Algebra through the development of \textit{Morpheus}\cite{morpheus}, a library for dynamic sparse matrices and algorithms.

\section{Conclusions}

In this work, we describe a design approach that allows the development of maintainable scientific software for heterogeneous computing architectures. The design is based on the abstract machine model that represents accurately the majority of HPC systems nowadays. The approach is based on meta-programming techniques using \texttt{C++} templates, which are used for the development of generic containers, routines and algorithms that serve as building-blocks for the development of performance-portable applications. We identify two concepts necessary for developing the generic building-blocks, the first, is the Memory Space that defines the location of the data, and the second is the Execution Space that distinguishes the implementation details. Our approach allows the integration of any programming model of choice as a back-end that provides the implementation details of the generic high-level application.  A FD scheme is developed as proof of concept, and benchmarked using roofline model analysis to demonstrate at least 90 \% performance efficiency on modern CPUs and GPUs. 

\section*{Acknowledgment}
This research is funded by Delphi Consortium at Delft University of Technology and the EPSRC project ASiMoV (EP/S005072/1). The experiments have been carried out on the Cyclone HPC system at the Cyprus Institute, and the Isambard~2 UK National Tier-2 HPC Service (http://gw4.ac.uk/isambard) operated by GW4 and the UK Met Office, and funded by EPSRC (EP/T022078/1).

%
%
%

\begin{thebibliography}{8}

\bibitem{host_device}
Ang J. A., et. al.: Abstract Machine Models and Proxy Architectures for Exascale Computing, Lawrence Berkeley National Laboratory, 2014.

\bibitem{raja}
Beckingsale D. A. et al.,: RAJA: Portable Performance for Large-Scale Scientific Applications. In: IEEE/ACM International Workshop on Performance, Portability and Productivity in HPC (P3HPC), Denver, CO, USA, 2019.


\bibitem{kokkos_i}
Carter H. et al.,: Kokkos: Enabling manycore performance portability through polymorphic memory access patterns. In: Journal of Parallel and Distributed Computing, Vol. 74, no. 12, pp. 3202-3216, 2014.


\bibitem{perf_port_across}
Deakin T. et al.,: Performance Portability across Diverse Computing Architectures. In: IEEE/ACM International Workshop on Performance, Portability and Productivity in HPC (P3HPC), Denver, CO, USA, 2019.

\bibitem{babel}
Deakin T. et al.,: Evaluating attainable memory bandwidth of parallel programming models via BabelStream. In: International Journal of Computational Science and Engineering. Special issue. Vol. 17, No. 3, pp. 247–262, 2018.

\bibitem{SoftArch}
Hasselbring W.: Software Architecture: Past, Present, Future. In: Gruhn V. Striemer R. (eds) The Essence of Software Engineering, Springer, pp. 169-184, 2018.

\bibitem{design}
Iglberger K.: C++ Software Design: Design Principles and Patterns for High-Quality Software, O’Reilly Media, Inc., 1005, 2022.

\bibitem{SEforCS}
Johanson N. A.: Software Engineering for Computational Science: Past, Present, Future. In: Computing in Science and Engineering, vol. 20, pp. 90-109, 2018.

\bibitem{meta}
Lilis Y. and Savidis A: A Survey of Metaprogramming Languages. In: ACM Computing Surveys, vol. 52, no. 6, pp. 1-39, 2019.

\bibitem{Prabhu}
Prabhu P. et al.,: A survey of the practice of computational science. In: Association for Computing Machinery, New York, NY, USA, Article 19, 1–12, 2011.

\bibitem{gometa}
Rompf T. et al.,: Go Meta! A case for Generative Programming and DSLs in Performance Critical Systems. In:  1st Summit on Advances in Programming Languages (SNAPL 2015), Asilomar, CA, USA, May 3-6, 2015.

\bibitem{stroustrou}
Stroustrup B.: The C++ Programming Language, Fourth Edition, ch. 17, pp. 481-526, Addison-Wesley, 2013.

\bibitem{trends}
Unat D. et al.,: Trends in Data Locality Abstractions for HPC Systems. In: IEEE Transactions on Parallel and Distributed Systems, vol. 28, no. 10, pp. 3007-3020, 1 Oct. 2017.

\bibitem{morpheus}
C. Stylianou and M. Weiland,: Exploiting dynamic sparse matrices for performance portable linear algebra operations. In 2022 IEEE/ACM International Workshop on Performance, Portability and Productivity in HPC (P3HPC). Los Alamitos, CA, USA: IEEE Computer Society, Nov 2022, pp. 47–57.

\end{thebibliography}
%

\end{document}